\documentclass[preprint,showpacs,preprintnumbers,amsmath,amssymb,nofootinbib]{revtex4}
\usepackage{graphicx}
\usepackage{subfigure}
\usepackage{dcolumn}
\usepackage{bm}

\def \D {{\cal D}}
\def \be {\begin{equation}}
\def \ee {\end{equation}}
\def \bea{\begin{eqnarray}}
\def \eea{\end{eqnarray}}

\begin{document}
\title{Time-Delay Interferometry with optical frequency comb}

\author{Massimo Tinto}
\email{Massimo.Tinto@jpl.nasa.gov}
\affiliation{Jet Propulsion Laboratory, California Institute of
  Technology, Pasadena, CA 91109}

\author{Nan Yu}
\email{Nan.Yu@jpl.nasa.gov}
\affiliation{Jet Propulsion Laboratory, California Institute of
  Technology, Pasadena, CA 91109}
           
\date{\today}

\begin{abstract}

  Heterodyne laser phase measurements in a space-based gravitational
  wave interferometer are degraded by the phase fluctuations of the
  onboard clocks, resulting in unacceptable sensitivity performance
  levels of the interferometric data. In order to calibrate out the
  clock phase noises it has been previously suggested that additional
  inter-spacecraft phase measurements must be performed by modulating
  the laser beams. This technique, however, considerably increases
  system complexity and probability of subsystem failure.

  With the advent of self-referenced optical frequency combs, it is
  possible to generate the heterodyne microwave signal that is
  coherently referenced to the onboard laser. We show in this case
  that the microwave noise can be cancelled directly by applying
  modified second-generation Time-Delay Interferometric combinations
  to the heterodyne phase measurements. This approach avoids use of
  modulated laser beams as well as the need of additional ultra-stable
  oscillator clocks.

\end{abstract}

\pacs{04.80.Nn, 95.55.Ym, 07.60.Ly}
\maketitle

\section{Introduction}
\label{Introduction}

Gravitational waves (GWs) are predicted by Einstein's theory of
general relativity and represent disturbances of space-time
propagating at the speed of light. Because of their extremely small
amplitudes and interaction cross-sections, GWs carry information about
regions of the Universe that would be otherwise unobtainable through
the electromagnetic spectrum.  Once detected, GWs will allow us to
open a new observational window to the Universe, and perform a unique
test of general relativity~\cite{Thorne}.

Since the first pioneering experiments by Joseph Weber in the early
sixties~\cite{Weber}, several experimental groups around the world
have been attempting to detect GWs. The coming on-line of
next-generation gravitational wave interferometers~\cite{LIGO,VIRGO},
however, is likely to break this sequence of experimental drawbacks
and result into the first detection before the end of the current
decade.

Contrary to ground-based detectors, which are sensitive to
gravitational waves in a band from about a few tens of Hz to a few
kilohertz, space-based interferometers are expected to access the
frequency region from a few tenths of millihertz to about a few tens
of Hz, where GW signals are expected to be larger in number and
characterized by larger amplitudes. The most notable example of a
space interferometer, which for several decades has been jointly
studied in Europe and in the United States of America, is the
Laser Interferometer Space Antenna (LISA) mission~\cite{LISA}.  By
relying on coherent laser beams exchanged among three remote spacecraft
along the arms of their forming giant (almost) equilateral triangle of
arm-length equal to $5 \times 10^6$ km, LISA aims to detect and study
cosmic gravitational waves in the $10^{-4} - 1$ Hz band.

A space-based laser interferometer detector of gravitational waves
measures relative frequency changes experienced by coherent laser
beams exchanged by three pairs of spacecraft.  As the laser beams are
received, they are made to interfere with the outgoing laser
light. Since the received and receiving frequencies of the laser beams
can be different by tens to perhaps hundreds of MHz (consequence of
the Doppler effect from the relative inter-spacecraft velocities and
the intrinsic frequency differences of the lasers), to remove these
large beat-notes present in the heterodyne measurements one relies on
the use of a microwave signal generated by an onboard clock (usually
referred to as Ultra-Stable Oscillator (USO)).  The magnitude of the
frequency fluctuations introduced by the USO into the heterodyne
measurements depends linearly on the USOs' noises themselves and the
heterodyne beat-note frequencies determined by the
inter-spacecraft relative velocities.  Space-qualified, state of the
art clocks are oven-stabilized crystals characterized by an Allan
deviation of ${\sigma_A} \approx 10^{-13}$ for averaging times of $1 -
1000$ s, covering most of the frequency band of interest to
space-based interferometers~\cite{LISA,eLISA,NGO}. In the case of the
LISA mission, in particular, it was estimated~\cite{TEA02} that the
magnitude of the power spectral density of the USO's relative
frequency fluctuations appearing, for instance, in the unequal-arm
Michelson Time-Delay Interferometry (TDI) combination $X$ would be
about six orders of magnitude larger than those due to the residual
(optical path and proof-mass) noise sources. 

A technique for removing the USO noise from the TDI combinations was
devised (see \cite{Hellingsetal, Hellings, TEA02} for more
details). This technique requires the modulation of the laser beams
exchanged by the spacecraft, and the further measurement of six more
inter-spacecraft relative phases by comparing the sidebands of the
received beam against sidebands of the transmitted beam. The physical
reason behind the use of modulated beams for calibrating the USOs
noises is to exchange the USOs phase fluctuations with the same time
delay as their lasers among the three spacecraft by performing
side-bands/side-bands measurements~\cite{Hellingsetal, Hellings,
  TEA02}. In so doing, additional six phase measurements are generated
that allow one to calibrate out the USOs phase fluctuations from the
TDI combinations while preserving the gravitational wave signal in the
resulting USO-calibrated TDI data.

It should be noticed, however, that if we could coherently transfer
the laser phase fluctuations to the microwave signal used in the
heterodyne measurements, then we would need to cancel only one noise
(the laser noise), which might be possible by deriving some new TDI
combinations. Coherently linking optical laser frequencies to
microwave frequencies has been thought to be impossible because of the
inability to directly count the optical frequency of a laser. However,
with the recent advent of the self-referenced octave-span optical
frequency comb (OFC) scheme (for which Hall~\cite{Hall} and
H\"ansch~\cite{Hansch} received the physics Nobel Prize in $2005$) it
is now possible to generate a microwave frequency signal that is
coherent to the frequency of the laser at a level significantly better
than the frequency stability required of a USO to avoid the
modulation-driven USO noise calibration procedure.

Although most of the reported implementations of the self-referenced
frequency comb technique have been performed in a laboratory
environment, active developments are being made for space-qualified
OFC systems~\cite{Holzwarth, PulseSpace}. In addition, more recent
micro comb source developments promise much smaller and integratable
comb devices~\cite{Kippenberg2011}. If each spacecraft could then
generate heterodyne measurements by relying on the OFC technique, then
the question to be answered is whether there exist new TDI
combinations that could account for the modified laser noise transfer
functions into the heterodyne measurements. The answer to this
question is yes, and in what follows we will derive their expressions.

In Section~\ref{SECII} we give a brief summary of the OFC technique,
and of characteristics of a self-referenced comb.  In
Section~\ref{SECIII} we then exemplify through a simple
interferometric configuration illustrating how the ``USO noise'' can
be canceled with TDI when the OFC technique is implemented.  In the
same section we then move on to the realistic configuration of a
space-based interferometer (such as LISA) and derive the expressions
for the twelve inter-spacecraft heterodyne phase measurements when the
OFC technique is used. This allows us to obtain the new
second-generation (i.e. flex-free) TDI
combinations~\cite{CH03,STEA03,TEA04,TD2005} that simultaneously
cancel the laser and the comb-generated microwave signal phase
noises. In Section~\ref{SECIV} we finally present our concluding
remarks, and emphasize that the use of the OFC technique will result
into a hardware and system design simplification, and increase system
reliability of future space-based gravitational wave interferometers.

\section{The optical frequency comb technique}
\label{SECII}

An OFC is a set of narrow spectral lines equally spaced in the
frequency domain. More importantly, when all the comb lines are phase
coherent (i.e. have fixed phase relationship) the superposition of all
the comb lines manifest themselves as a periodical wave train in the
time domain. The resulting repetition rate is equal to the comb
spacing and it is governed by the Fourier transform relationship. More
often, the comb lines are in phase so the time domain waveform is a
train of short pulses separated by the inverse of the repetition
rate. The pulse width is then Fourier transform limited by the comb
spectral width.

\begin{figure}
\centering
\includegraphics[width = 6in]{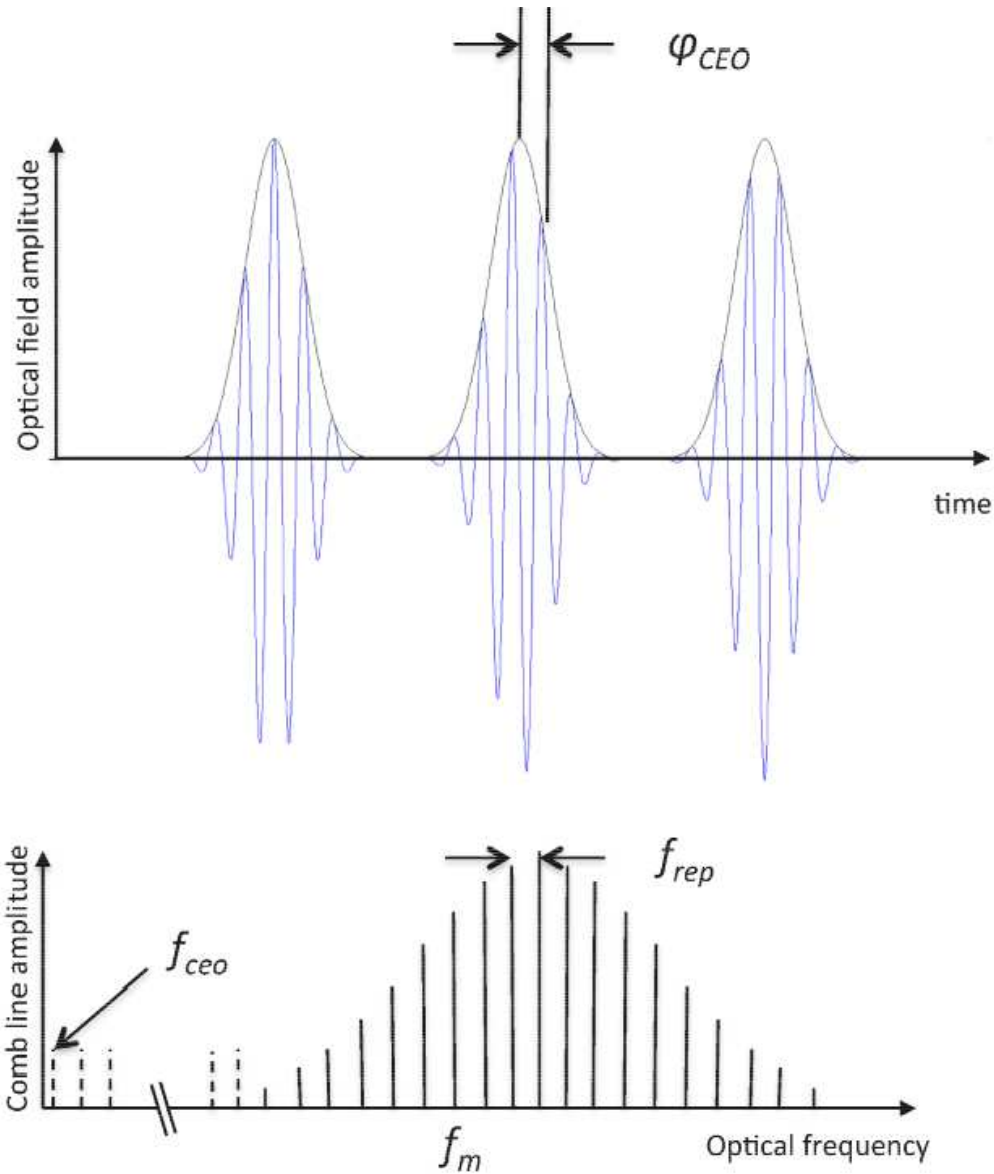}
\caption{Illustration of a mode locked laser pulse train and its
  corresponding comb in the frequency domain. The top plot in
  the time domain shows the carrier envelope offset phase and its
  changes. The bottom plot shows the comb lines and the carrier
  envelope offset frequency near the DC frequency.}
\label{Pulsetrain}
\end{figure}

Typical OFCs are generated by mode-locked lasers in which multiple
modes are locked in phase, outputting short pulse trains. The pulse
train propagates with a group velocity while the underline optical
carrier travels at the phase velocity. When the modes are exactly in
phase, the peak of the carrier is aligned with the peak of the pulse
envelope, as shown in the first pulse in figure ~\ref{Pulsetrain}. In
the absence of any dispersion, this relationship holds over time as
the pulse propagates. In practice,
dispersion changes the relative phases of the modes, and the peak of
the carrier amplitude shifts from that of the pulse envelope. This
offset is typically referred to as Carrier Envelope Offset (CEO) phase
shift and the rate of this offset phase change over a period gives the
CEO frequency~\cite{combbook}, as shown in the frequency domain
spectrum in figure~\ref{Pulsetrain}. In particular, this implies the well
known mode frequency relation: $f_m = f_{ceo}+ m f_{rep}$. Since
$f_{ceo}$ is affected by dispersion, as there are many environmental
effects influencing it, $f_{ceo}$ is not a stable parameter and, as a
consequence, there is no fixed frequency or phase relationship between
the optical carrier and the pulse envelope, i.e. the beat-note of the
rf signals among the modes. For many years before the recent
self-referenced OFC development, the two frequencies were kept
separate and used independently.

Recent advancements in the study of ultra-fast phenomena and the field
of optical frequency metrology have revolutionized the use of OFC with
abilities to measure and control the CEO. As shown in reference
~\cite{combbook}, if one compares the $N_{m}$th harmonic of the $m$th
mode with the $N_{m'}$th harmonic of the $m'$th mode, the beat
note of the two yields $(N_m - N_{m'})f_{ceo}$. In particular, if the
comb has an octave spanning width where one can choose $N_m=2$ and
$N_{m'}=1$, the mixing beat-note of 1f-2f directly gives the measure
of the CEO frequency. Like in a typical phase locking loop, the measured
CEO phase can be used to stabilize the CEO itself, resulting in an OFC
that has the well-defined relationship between the optical frequencies
$f_m$ in the modes and the rf frequency of the repetition rate
$f_{rep}$, now referred to as {\it
  self-referencing}~\cite{selfreferencing}. Such control capability
not only allows one to align carrier amplitude peaks with the short
pulse envelopes in ultrafast phenomena study, but more revolutionarily
provides the optical frequency divider ability down to microwave in
one simple step. In the latter case with a self-referenced comb, what
it means is that the optical field in the comb has a coherent phase
relationship with its rf beat-note signal and one can count the
optical frequency by counting the rf frequency. Furthermore, the
optical phase noise is also directly down converted to the rf phase
noise. Recent studies have shown that the comb frequency precision can
be at $1 \times 10^{-19}$ ~\cite{Comb} while the CEO phase can be
controlled down to the mrad level. In the context of the investigation
in this paper, the properties of the phase coherent optical divider
are exploited for the onboard clock phase noise cancellation and for
the generation of stable onboard rf signals directly from the comb.

The most mature self-referenced OFCs are generated with a combination
of femtosecond mode locked laser and super-continuum generation to
achieve octave spanning spectrum width. While both Ti:Saphire and
fiber mode locked lasers are successfully implemented for self
referencing, the most suitable laser source for space interferometer
mission may be fiber laser based systems. In fact, self-referencing
comb systems are now commercially available, mostly for laboratory
applications. Efforts are being made in developing the fiber OFC
technology for space applications. A flight OFC system is being
developed for sounding rocket experiments~\cite{Holzwarth}, and a
femtosecond fiber laser has already been flown in
space~\cite{PulseSpace}. It is in fact expected that a flight OFC
system, suitable for space missions, will become available in the near
future.

More recently it has been shown that coherent OFC can also be
generated using high-Q micro resonators through Kerr nonlinear four
wave mixing process~\cite{microcomb}. In this case, a cw laser is
resonantly injected into the resonator and strongly confined in the
small volume mode. The high Q-factor leads further field buildup in
the resonator, which enhances the Kerr nonlinear effect in the micro
resonator. Various cavity eigenmodes can, therefore, be excited
through four-wave mixing as the laser light sequentially cascades from
the pump to the other modes~\cite{chembo}, forming an optical comb
with spacing determined by the resonator mode space, free spectral
range. When the condition is right, these modes are phase locked in a
soliton form, very much like a typical mode locked
laser~\cite{soliton}. In comparison with their mode-locked lasers
counterparts, however, these whispering galery mode optical frequency
comb generators are characterized by a significantly reduced size and
power consumption, along with a high repetition rate. They are,
therefore, particularly suitable for miniaturization, chip
integration, and space applications. For the purpose of
self-referencing in optical metrology, it is important that the comb
spans over one octave for 1f-2f self-reference, or at least 2/3 of an
octave for 2f-3f self-referencing.  While self-referenced micro comb
are still been developed in research labs, we believe space-qualified,
compact Kerr combs will be available in the near future.

The striking property of the self-referenced OFC is the phase
coherence between the optical carriers and their rf beat note
signal. As it will be shown later, this allows the laser and resulting
microwave signal phase noises cancellation elegantly in the
generalized TDI combinations. Furthermore, one can take advantage of
the onboard frequency stabilized laser system to generate a stable
microwave signal. Since the onboard stabilized lasers are expected to
achieve a stability at the level of $10^{-14}/\sqrt{Hz}$ or better over the frequency range of interests to space
interferometer gravitational wave detection,
by phase-locking the optical frequency comb to the stabilized laser,
the corresponding rf beat-note stability will be also of a few parts
in $ 10^{14}$. Therefore, the stabilized
OFC technique will provide an rf signal that can be significantly more 
stable of the current best space-qualified USOs.  The OFC
implementation will not only eliminate the need for separate USOs but 
also remove entirely the need for modulated beams used by the USO 
calibration method, as it will be shown in the following sections.

\section{Time-Delay Interferometry with optical frequency comb}
\label{SECIII}

In the following subsections we derive the modified expressions of the
second-generation~\cite{CH03,TEA04,TD2005} TDI combinations that
simultaneously cancel the laser and the microwave phase noises when
the OFC technique is implemented. To physically understand how this is
possible, we will first consider a simple interferometer configuration
in which the rate-of-changes of the arm-lengths are large enough to
require the removal of the beat-notes from the two-way data but small
enough to treat the arm-lengths as constant.  This will allow us to
establish the basis for then deriving new second-generation TDI
combinations that a three-arms, six laser links space-based
interferometer can generate.

\subsection{TDI with OFC: an example application}
\label{SECIIIa}
 
\begin{figure}
\centering
\includegraphics[width = 6in]{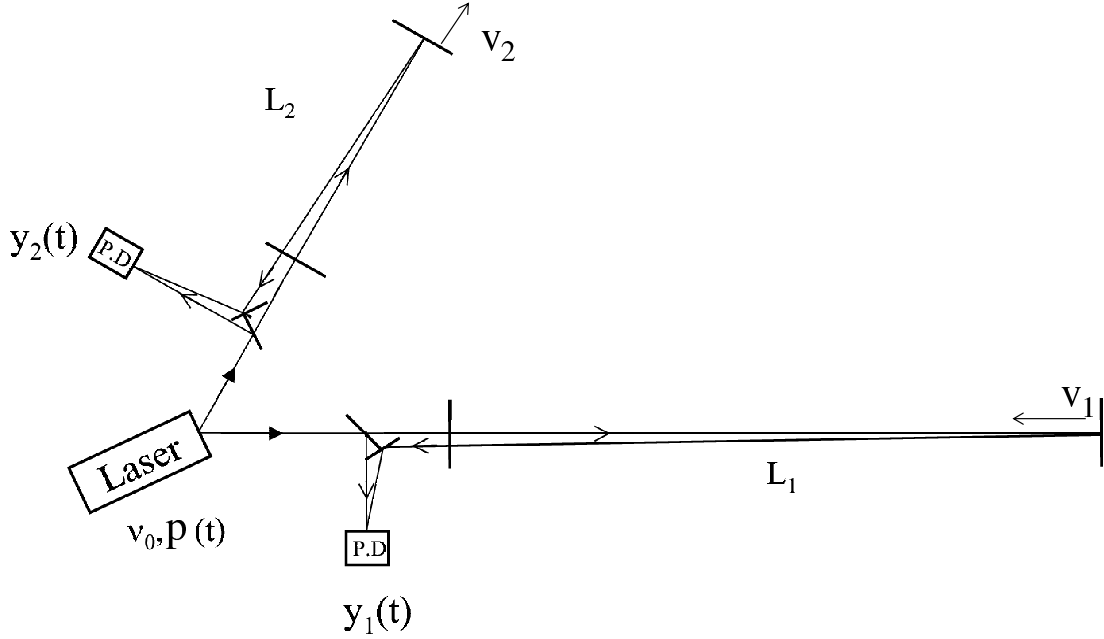}
\caption{Light from a laser is split into two beams, each injected
  into an arm formed by pairs of mirrors moving with velocities $V_1$
  and $V_2$ relative to the inertial reference frame with respect to
  which the laser and the beam-splitter are at rest. Since the length
  of the two arms, $L_1$ and $L_2$, are different, the light beams
  from the two arms are not recombined at one photo detector. Instead
  each is separately made to interfere with the light that is injected
  into the arms. Two distinct photo detectors are now used, and phase
  (or frequency) fluctuations are then monitored and recorded
  there. See text for details.}
\label{2unequalarms}
\end{figure}

Let us consider the following idealized unequal-arm interferometer
whose end-mirrors are moving with velocities $V_1$ and $V_2$ along the
directions of arms 1 and 2 respectively (see figure
\ref{2unequalarms}). These velocities are relative to the inertial
reference frame in which the laser and the beam-splitters are at
rest. The two light beams coming out of the two arms are not made to
interfere at a common photodetector as each is made to interfere with
the incoming light from the laser at a photodetector. This is because
direct recombination of the beams at a single photodetector would not
suppress the laser phase noise in the phase difference measurement due
to the inequality of the arms. \footnote{By using two independent
  photodetectors we decouple the phase fluctuations experienced by the
  two beams in the two arms and obtain two (rather than one)
  ``two-way'' measurements. This ``data redundancy'' allows us to
  cancel the laser phase noise in the data while retaining a possibly
  present gravitational wave signal. This is done by properly
  time-shifting and linearly combining in digital form the two two-way
  phase differences.}

The velocities of the mirrors can be assumed to have frequency
components that are outside the band of operation of the
interferometer, while their amplitudes are such as to require the use
of a numerically controlled oscillator (NCO) (driven by a clock) to
remove the beat-notes in the two ``two-way'' heterodyne phase
measurements, $\phi_1(t)$ and $\phi_2(t)$. As a result of this
``down-conversion'' operation, the two-way data assume the following
forms \footnote{Throughout this article the speed of light $c$ and $2
  \pi$ are assumed to be equal to $1$}

\begin{eqnarray}
\phi_1(t) & = & [\nu_0 (1 - 2V_1) - \nu_0 - a_1 f]t + p(t - 2L_1) -
p(t) - a_1 q(t) + h_1(t) + n_1(t),
\label{2Doppler1}
\\
\phi_2(t) & = & [\nu_0 (1 - 2V_2) - \nu_0 - a_2 f]t + p(t - 2L_2) -
p(t) - a_2 q(t) + h_2(t) + n_2(t),
\label{2Doppler2}
\end{eqnarray}
where $a_1$, $a_2$ are selected by the NCO in such a way to suppress
the beat-notes from the two phase measurements to a required value,
$f$ is the microwave frequency generated by the clock driving the
down-conversion, $h_1$, $h_2$ are the contributions from a possibly
present gravitational wave signal, and $n_1$, $n_2$ are the random
processes associated to the overall remaining noises affecting the two
phase measurements. Note that in Eqs. (\ref{2Doppler1},
\ref{2Doppler2}) we have denoted with $q(t)$ the random process
associated with the phase noise from the clock driving the NCO
~\cite{TEA02} and $p(t)$ the laser noise. Since the coefficients $a_i \ , \ i=1, 2$ can be made
to be equal to $a_i = - 2V_i \nu_0/f \ , \ i = 1, 2$, then 
Eqs. (\ref{2Doppler1}, \ref{2Doppler2}) become
\begin{eqnarray}
\phi_1(t) & = & p(t - 2L_1) - p(t) - a_1 q(t) + h_1(t) + n_1(t) \ ,
\label{2Doppler12}
\\
\phi_2(t) & = & p(t - 2L_2) - p(t) - a_2 q(t) + h_2(t) + n_2(t) \ ,
\label{2Doppler22}
\end{eqnarray}
If we now assume the clock frequency $f$ to be the result of
implementing a self-referenced OFC that is driven by the onboard laser, then the
following relationship between the phase fluctuations of the local
oscillator, $q(t)$, and those of the laser, $p(t)$, holds
\begin{equation}
q (t) = \frac{f}{\nu_0} \ p(t) \ + \Delta q(t) \ .
\label{q1}
\end{equation}
In Eq. (\ref{q}) we have denoted with $\Delta q(t)$ the residual noise
representing the level of coherence between the frequency of the laser
and the microwave frequency referenced to it. It has been demonstrated
experimentally ~\cite{Comb} that the magnitude of this residual phase
noise can be reduced to a level corresponding to an Allan standard
deviation of a few parts in $10^{-19}$ over time scales of interest to
space-based GW interferometers. Since this value is three orders of
magnitude smaller than the USO noise level required for avoiding the
use of the calibration procedure based on modulated
beams~\cite{TEA02}, we can then safely disregard it in our analysis.
After substituting Eq. (\ref{q1}) into Eqs. (\ref{2Doppler12},
\ref{2Doppler22}) we obtain the expressions for the two-way phase
differences when the OFC technique is implemented
\begin{eqnarray}
\phi_1(t) & = & p(t - 2L_1) - p(t) + 2V_1 p(t) + h_1(t) + n_1(t) \ ,
\label{2Doppler13}
\\
\phi_2(t) & = & p(t - 2L_2) - p(t) + 2V_2 p(t) + h_2(t) + n_2(t) \ .
\label{2Doppler23}
\end{eqnarray}
Following ~\cite{TA98} it is then straightforward to show that the
following combination of the two two-way phase measurements cancels
the laser noise
\begin{equation}
  X^{OFC} (t) \equiv [\phi_1(t - 2L_2) - (1 - 2V_2) \phi_1(t)] - 
  [\phi_2(t - 2L_1) - (1 - 2V_1) \phi_2(t)] \ .
\label{XOFC}
\end{equation}
In other words, all $p(t)$ and $q(t)$ terms (through Eq.~\ref{q1})
drop out in $ X^{OFC}$. Although the above expression of the
unequal-arm Michelson TDI combination reflects the assumption that the
velocities of the mirrors are large enough to require the removal of
the beat-notes from the two-way data but small enough to treat the
arm-lengths as constant, it shows that the implementation of the OFC
technique allows us to simultaneously cancel the laser and microwave
noises with TDI.  In the following sections we will remove the
assumption of constant arm-lengths and derive the TDI expressions that
account for the rotation of the array (Sagnac effect) as well as the
time-dependence (``flexi'') of its arms~\cite{CH03,TEA04,TD2005}.

\subsection{Generalized TDI formulation}
\label{SECIIIb}

In order to derive the new TDI combinations valid when the OFC
technique is used for generating a microwave signal coherent to the
onboard laser, we will follow the description of a space-based
interferometer (such as LISA and eLISA/NGO with its three operational
arms ~\cite{LISA,eLISA,NGO}) adopted in earlier publications (see
reference ~\cite{TD2005} and references therein).

\begin{figure}
\includegraphics[width=3.0in, angle=0]{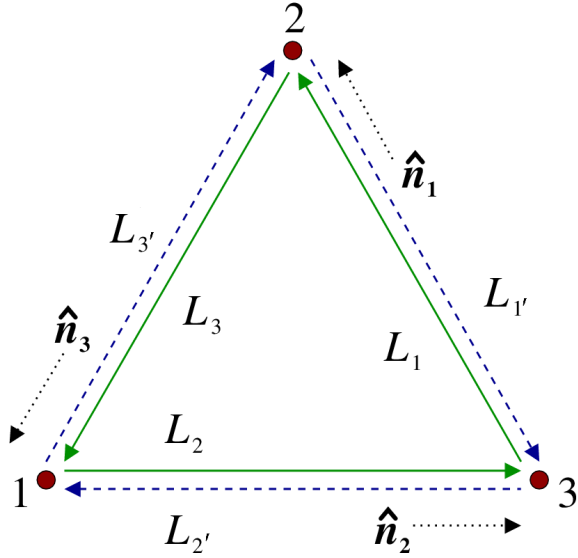}
\caption{Schematic array configuration. The spacecraft are labeled
      1, 2, and 3. The optical paths are denoted by $L_i$, $L_{i'}$
      where the index $i$ corresponds to the opposite spacecraft. The
      unit vectors $\hat{{\bf n}}_i$ point between pairs of
      spacecraft, with the orientation indicated.}
\label{fig1}
\end{figure}

As shown in Figure~\ref{fig1}, where the overall geometry of the
interferometer is defined, there are three spacecraft, six optical
benches, six lasers, six proof-masses, and twelve photo
detectors. There are also six phase difference data going clock-wise
and counter-clock-wise around the triangle. The spacecraft are labeled
$1, 2, 3$ and the optical paths traveled by the light beams as they
propagate along the arms of the constellation are denoted $L_i$ and
$L_{i'}$, with $i = 1, 2, 3$ being opposite spacecraft $i$. The two
optical paths along the same arm are different because the rotational
motion of the array results into a difference of the light travel
times in the two directions around a Sagnac circuit~\cite{S03,
  CH03}. As shown in Figure~\ref{fig1} we have conventionally denoted
with $L_{i'}$ and $L_i$ the optical delays for clockwise and
counter-clock-wise propagation respectively. Furthermore, since $L_i$
and $L_{i'}$ not only differ from one another but can be time
dependent (they ``flex''), new second-generation TDI combinations were
derived in order to account for this effect. This was done by using
the non-commuting time-delay operators formalism discussed
in~\cite{TD2005}, and we will rely on it for the derivation of the new
second-generation TDI combinations that simultaneously cancel the
noises of the microwave signals generated by the OFC technique.

In Figure~\ref{fig1} the vertices 1, 2, 3 are oriented clockwise,
while the unit vectors between spacecraft are denoted $\hat n_i$,
oriented as indicated in the figure. We index the phase difference
data to be analyzed as follows: the beam arriving at spacecraft $i$
has subscript $i$ and is primed or unprimed depending on whether the
beam is traveling clockwise or counter-clockwise (the sense defined
here with reference to Figure~\ref{fig1}) around the triangle,
respectively. Thus, as seen from the figure, $s_{1}$ is the phase
difference time series measured at reception at spacecraft~1 with
transmission from spacecraft~2 (along $L_3$).

Similarly, $s_{1'}$ is the phase difference series derived from
reception at spacecraft~1 with transmission from spacecraft~3. The
other four one-way phase difference time series from signals exchanged
between the spacecraft are obtained by cyclic permutation of the
indices: $1 \to 2 \to 3 \to 1$. We also adopt the notation for delayed
data streams described in~\cite{TD2005}, in which there are six
time-delay operators $\D_i \ , \D_{i'}$, $i = 1, 2, 3$, and where, for any
data stream $\Psi(t)$,
\begin{equation}
  \D_i \Psi(t) = \Psi(t - L_i) \ , \ \D_{i'} \Psi(t) = \Psi(t - L_{i'}) \ .
\end{equation}
For example, $\D_2 s_{1} (t) = s_{1}(t - L_2)$, $\D_{2'}
\D_{3'} s_{1}(t) = s_{1}(t - L_{3'}(t - L_{2'}) - L_{2'} (t))$,
etc. Note that the operators do not commute, as: $\D_{3'} \D_{2'}
s_{1}(t) = s_{1}(t - L_{2'}(t - L_{3'}) - L_{3'} (t)) \neq \D_{2'}
\D_{3'} s_{1}(t) = s_{1}(t - L_{3'}(t - L_{2'}) - L_{2'} (t))$.

Following~\cite{TEA02}, the heterodyne phase measurements, $s_i (t) \ , \ s_{i'} \ , \
i=1, 2, 3$ can be written in the following form

\begin{eqnarray}
s_1(t) &=& [\nu_{2'} (1 - {\dot{L_3}}) - \nu_1 - a_1 f_1]t + \D_3
  p_{2'} - p_1(t) - a_1 q_1(t) + s_1^{gw}(t) + N_1 (t) \ ,
\nonumber
\\
s_{1'}(t) &=& [\nu_3 (1 - {\dot{L_{2'}}}) - \nu_{1'} - a_{1'} f_{1'}]t + 
\D_{2'} p_3 - p_{1'}(t) - a_{1'} q_{1'}(t) + s_{1'}^{gw}(t) + N_{1'} (t) \ ,
\label{S}
\end{eqnarray}
where we have denoted with $p_i$, $p_{i'}$ the lasers phase
fluctuations, the microwave frequencies $f_1 \ , f_{1'}$ are generated
by the OFCs driven by lasers $1$ and $1'$ respectively, and the
coefficients $a_1$, $a_{1'}$ are synthesized by a
numerically-controlled oscillator (NCO) to be equal to the following
values~\cite{TEA02}
\begin{equation}
a_1 = \frac{\nu_{2'} (1 - {\dot{L_3}}) - \nu_1}{f_1} \ , \ 
a_{1'} = \frac{\nu_3 (1 - {\dot{L_{2'}}}) - \nu_{1'}}{f_{1'}} \ .
\label{a_coeff}
\end{equation}
In Eq. (\ref{S}) we have denoted with $N_1 \ , \ N_{1'}$ all the
remaining noises affecting the one-way measurements, and with
$s_{1}^{gw} \ , \ s_{1'}^{gw}$ the contribution to the one-way Doppler
measurements from a passing gravitational wave signal.

Consistently with earlier work~\cite{TEA02}, Eqs. (\ref{S}) reflect
the assumption of having the frequencies of the lasers different from
each other.  Note that the phase fluctuations from the microwave
frequency references driving the NCO enter into $s_1$ and $s_{1'}$
through the terms $a_1 q_1$ and $a_{1'} q_{1'}$ respectively. Since
these phase fluctuations associated to the microwave signals have been
generated by the frequency comb subsystems, they are related to the
laser phase fluctuations, $p_1$ and $p_{1'}$, through the following
relationships
\begin{equation}
q_1 (t) = \frac{f_1}{\nu_1} p_1 (t) \ + \Delta q_1 (t) \ , \ 
q_{1'} (t) = \frac{f_{1'}}{\nu_{1'}} p_{1'} (t) \ + \Delta q_{1'} (t) \ .
\label{q}
\end{equation}
In Eq. (\ref{q}) we have denoted with $\Delta q_1 (t)$, $\Delta q_{1'}
(t)$ the residual noises representing the level of coherence between
the frequencies of the lasers and the microwave frequencies referenced
to them. The magnitude of this residual phase noise has been
demonstrated experimentally ~\cite{Comb} to correspond to an Allan
standard deviation of a few parts in $10^{-19}$ over time scales of
interest to space-based GW interferometers. Since this is three orders
of magnitude smaller than the USO noise required for avoiding the use
of the calibration procedure based on modulated beams~\cite{TEA02}, we
infer that the contribution from the noises $\Delta q_i \ , \ \Delta
q_{i'} \ , \ i=1, 2, 3$ to the overall noise budget of the TDI
combinations can be regarded as negligible. For this reason from now
on they will be disregarded in our analysis.

If we now substitute Eqs.(\ref{a_coeff}, \ref{q}) into Eq.(\ref{S}),
the heterodyne measurements $s_1$, $s_{1'}$ assume the following forms
\begin{eqnarray}
s_1(t) &=& \D_3 p_{2'} - p_1(t) - A_1 p_1(t) + s_1^{gw}(t) + N_1 (t) \ ,
\nonumber
\\
s_{1'}(t) &=& \D_{2'} p_3 - p_{1'}(t) - A_{1'} p_{1'}(t) + s_{1'}^{gw}(t) + N_{1'} (t) \ ,
\label{Snew}
\end{eqnarray}
where the coefficient $A_1 \ , \ A_{1'}$ are equal to 
\begin{equation}
A_1 \equiv \frac{a_1 f_1}{\nu_1} = \frac{\nu_{2'} (1 - {\dot{L_3}}) - \nu_1}{\nu_1} \ , \ 
A_{1'} \equiv \frac{a_{1'} f_{1'}}{\nu_{1'}} = 
\frac{\nu_3 (1 - {\dot{L_{2'}}}) - \nu_{1'}}{\nu_{1'}} \ .
\end{equation}
Note that, over periods of a few light-times, the coefficients $A_i$,
$A_{i'}$ ($i = 1, 2, 3$) can be regarded as constant since the
spacecraft relative velocities start to display a noticeable change
only over periods of weeks to months~\cite{Folkner}.

\begin{figure}
\includegraphics[width=4.0in, angle=0]{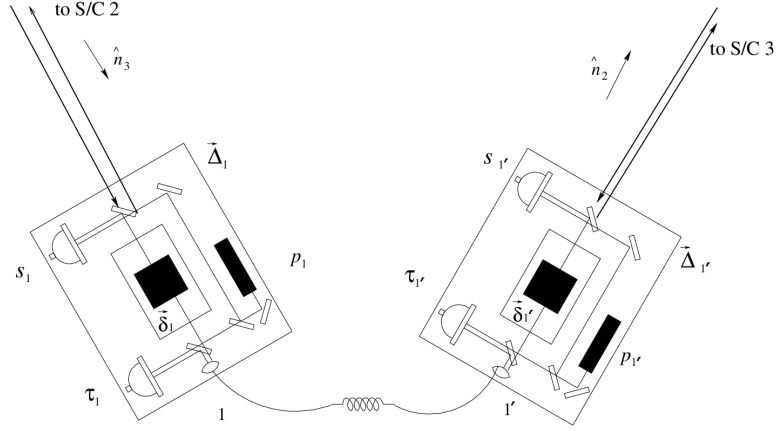}
\caption{Schematic diagram of proof-masses-plus-optical-benches
      onboard the spacecraft. The left-hand bench reads out the phase
      signals $s_{1}$ and $\tau_{1}$. The right-hand bench analogously
      reads out $s_{1'}$ and $\tau_{1'}$. The random displacements of
      the two proof masses and two optical benches are indicated (lower
      case $\vec \delta_{i} , \vec \delta_{i'}$ for the proof masses,
      upper case $\vec \Delta_{i} , \Delta_{i'}$ for the optical
      benches).}
\label{fig2}
\end{figure}

If we assume the optical bench design to be as shown in Figure
\ref{fig2}, then six more phase difference series result from laser
beams exchanged between adjacent optical benches within each
spacecraft; these are similarly indexed as $\tau_{i}$, $\tau_{i'}$, $i
= 1, 2, 3$. \footnote{The optical bench design shown in
  figure~\ref{fig2} represents one of the possible configurations for
  integrating the onboard drag-free system with the heterodyne
  measurements. Although other optical bench designs result into
  different inter-proof-mass phase measurements~\cite{eLISA,MGRS}, the
  resulting expressions for the new TDI combinations derived in this
  article are all equivalent.} The photo receivers that generate the
data $s_{1}$, $s_{1'}$, $\tau_{1}$, and $\tau_{1'}$ at spacecraft~1
are also shown in figure~\ref{fig2}.

The adopted optical bench design shows optical fibers transmitting
signals both ways between adjacent benches. We ignore time-delay
effects for these signals and will simply denote by $\mu_i(t)$ the
phase fluctuations upon transmission through the fibers of the laser
beams with frequencies $\nu_{i}$, and $\nu_{i'}$. The $\mu_i (t)$
phase shifts within a given spacecraft might not be the same for large
frequency differences $\nu_{i} - \nu_{i'}$. For the envisioned
frequency differences (a few hundred MHz), however, the remaining
fluctuations due to the optical fiber can be assumed to be independent
of the direction of light propagation through them~\cite{ETA00}.

If we limit our attention only to the terms containing the laser phase
noises, we can then write down the following expressions for the
$\tau_{i}$, $\tau_{i'}$ phase measurements
\begin{eqnarray}
\tau_{1} (t) & = & [\nu_{1'} - \nu_1 - c_1 f_1]t + p_{1'}(t) - p_{1}(t)
- c_1 \frac{f_{1}}{\nu_{1}} p_1(t) \ ,
\label{eq:due}
\\
\tau_{1'}(t) & = &  [\nu_1 - \nu_{1'} - c_{1'} f_{1'}]t + p_{1}(t) - p_{1'}(t)
- c_{1'} \frac{f_{1'}}{\nu_{1'}} p_{1'}(t) \ ,
\label{eq:quattro}
\end{eqnarray}
where, like in the case of the inter-spacecraft measurements $s_1$,
$s_{1'}$, the coefficient $c_1$, $c_{1'}$ are determined by the NCO to
be equal to
\begin{equation}
c_1 = \frac{\nu_{1'} - \nu_1}{f_1} \ , \ c_{1'} = \frac{\nu_1 -
  \nu_{1'}}{f_{1'}} \ .
\label{c_coeff}
\end{equation}
Note that in Eq. (\ref{c_coeff}) we have used the relationship between
the phase noise of the microwave signal and that of its driving laser
(Eq. (\ref{q})). As discussed in other TDI
references~\cite{TEA02,TD2005} the inter-bench phase measurements
$\tau_i$, $\tau_{i'}$ enter into the TDI combinations in the form of
their differences.  It is therefore convenient to rewrite the set of
heterodyne measurements in the following form (for simplicity of
notation we will omit the contribution from the gravitational wave
signal and the other noises in the $s_1$ and $s_{1'}$ measurements)
\begin{eqnarray}
s_1 & = & \D_3 p_{2'} - (1 + A_1) p_1 \ ,
\label{s1}
\\
s_{1'} & = & \D_{2'} p_3 - (1 + A_{1'}) p_{1'} \ ,
\label{s1'}
\\
z_1 & \equiv & \frac{\tau_1 - \tau_{1'}}{2} = (1 + \frac{\rho_{1'}}{2}) p_{1'} - (1 +
\frac{\rho_1}{2}) p_{1} \ ,
\label{z1}
\end{eqnarray}
where $\rho_{1}$, $\rho_{1'}$, are equal to
\begin{equation}
\rho_1 \equiv \frac{\nu_{1'} - \nu_1}{\nu_1} \ \ , \ \ \rho_{1'} = 
 \frac{\nu_{1} - \nu_{1'}}{\nu_{1'}} \ .
\label{rho1}
\end{equation}
Six other relations, for the readouts at vertices 2 and 3, are given
by cyclic permutation of the indices in Eqs.~(\ref{s1}, \ref{s1'}, \ref{z1}).

In order to derive the new TDI combinations that rely on the OFC
technique, we will start by taking specific combinations of the
one-way data entering in each of the expressions derived
above. Following the approach in ~\cite{TEA04,TD2005}, these
combinations are identified by requiring the laser noises entering
into them to be only of one ``kind'', i.e. with primed or unprimed
indeces. After some simple algebra to account for the presence of the
heterodyne coefficients, these combinations of the one-way heterodyne
measurements (called $\eta_i \ , \ \eta_{i'}$ in ~\cite{TD2005}) here
assume the following forms

\begin{eqnarray}
\eta_{1} & \equiv & \frac{1 + \frac{\rho_{2'}}{2}}{1 +
  \frac{\rho_{2}}{2}} \ s_1 - \frac{D_3 z_2}{1 + \frac{\rho_{2}}{2}} = 
\D_3 p_2 - I_1 p_1 \ ,
\label{eta1}
\\
\eta_{1'} & \equiv & s_{1'} + \frac{1 + A_{1'}}{1 +
  \frac{\rho_{1'}}{2}} z_1 = \D_{2'} p_3 - I_{1'} p_1 \ ,
\label{eta2}
\end{eqnarray}
where the coefficients $I_1 \ , \ I_{1'}$ are given by the following expressions
\begin{equation}
I_1 \equiv \frac{(1 + A_{1'})(1 + \frac{\rho_{1}}{2})}{1 +
  \frac{\rho_{1'}}{2}}  \ , \qquad 
I_{1'} \equiv \frac{(1 + A_1)(1 + \frac{\rho_{2'}}{2})}{1 +
  \frac{\rho_{2}}{2}} \ ,
\label{I}
\end{equation}
and the other $\eta$-combinations are obtained as usual by
permutations of the spacecraft indeces.

The new TDI combinations are chosen in such a way so as to retain only
one of the three noises $p_i$, $i=1, 2, 3$, if possible. In this way
we can then implement an iterative procedure based on the use of these
basic combinations and of time-delay operators, to cancel the laser
noises after dropping terms that are quadratic in $\dot{L}/c$ or
linear in the accelerations~\cite{CH03,TEA04,TD2005}. This iterative
time-delay method, to first order in the velocity, is illustrated
abstractly as follows. Given a function of time $\Psi = \Psi(t)$, time
delay by $L_i$ is now denoted either with the standard comma
notation~\cite{AET99} or by applying the delay operator $\D_{i}$,
\begin{equation}
  \D_{i} \Psi = \Psi_{,i} \equiv \Psi(t - L_i(t)).
  \label{eq:25}
\end{equation}
We then impose a second time delay $L_j(t)$:
\begin{eqnarray}
  \D_{j} \D_{i} \Psi = \Psi_{;ij} & \equiv &  \Psi(t - L_j(t) - L_i(t - L_j(t)))
  \nonumber \\
  & \simeq  & \Psi(t - L_j(t) - L_i(t) + \dot  L_i(t) L_j)
  \nonumber \\
  & \simeq & \Psi_{,ij} + \dot \Psi_{,ij} \dot L_i L_j.
  \label{eq:26}
\end{eqnarray}
A third time delay $L_k(t)$ gives
\begin{eqnarray}
  \D_{k} \D_{j} \D_{i} \Psi = \Psi_{;ijk} & = &
  \Psi(t - L_k(t) - L_j(t - L_k(t)) - L_i(t - L_k(t) - L_j(t - L_k(t))))
  \nonumber \\
  &\simeq& \Psi_{,ijk} + \dot \Psi_{,ijk}
  \left[\dot L_i (L_j + L_k) + \dot L_j L_k\right],
  \label{eq:27}
\end{eqnarray}
and so on, recursively; each delay generates a first-order correction
proportional to its rate of change times the sum of all delays coming
after it in the subscripts. Commas have now been replaced with
semicolons~\cite{STEA03}, to remind us that we consider moving arrays.
When the sum of these corrections to the terms of a data combination
vanishes, the combination is called flex-free. Finally, it should be
noticed that each delay operator, $\D_i$ has a unique inverse
operator, $\D_i^{-1}$, whose expression can be derived by requiring
that $\D_i^{-1} \D_i = I$ and neglecting quadratic and higher-order
velocity terms. Its action to a time series $\Psi(t)$ is
\begin{equation}
\D_i^{-1} \Psi(t) \equiv \Psi(t + L_i(t + L_i)) \ ,
\end{equation}
i.e. it advances the time-series by a delay $L_i$ estimated not at time $t$
but rather at time $t + L_i$.

%%%%%%%%%%%%%%%%%%%%%%%%%%%%%%%%%%%%%%%%%%%%%%%%%%%%%%%%%%%%%%%%%%%%%%%%%%%%%%%%%%%
%%%%%%%%%%%%%%%%%%%%%%%%%%%%%%%%%%%%%%%%%%%%%%%%%%%%%%%%%%%%%%%%%%%%%%%%%%%%%%%%%%%

\subsection{The unequal-arm Michelson}
\label{sec:michelson}

The unequal-arm Michelson combination synthesized onboard spacecraft
\# 1 relies on the four measurements $\eta_{1}$, $\eta_{1'}$,
$\eta_{2'}$, and $\eta_{3}$.  From Eqs. (\ref{eta1}, \ref{eta2}), after
some simple algebra, it is easy to show that the following two
combinations, $I_{2'} \eta_{1} + D_3 \eta_{2'}$ and $I_3 \eta_{1'} +
D_{2'} \eta_{3}$, contain only the noise from laser \# 1 and have
the following forms

\begin{eqnarray}
I_{2'} \eta_{1} + \eta_{2';3} & = & \left( \D_{3}\D_{3'} - I_1 I_{2'}
  \right) p_1 \ ,
\label{eq:diecia}
\\
I_3 \eta_{1'} + \eta_{3;2'}  & = & \left( \D_{2'}\D_{2} - I_{1'} I_3
\right) p_1 \ .
  \label{eq:diecib}
\end{eqnarray}
Since in the quasi-stationary case any pairs of these operators
commute, i.e., $\D_i \D_{j'} - \D_{j'} \D_i \approx 0$, from
Eqs.~(\ref{eq:diecia} and \ref{eq:diecib}) it is easy to derive the
following expression for the unequal-arm interferometric combination
$X^{OFC}$ that eliminates $p_1$:
\begin{equation}
  X^{OFC} = \left[\D_{2'}\D_{2} - I_{1'} I_3 \right] (I_{2'} \eta_{1} + \eta_{2';3}) -
  \left[\left(\D_{3}\D_{3'} - I_1I_{2'}\right)\right] (I_3 \eta_{1'} + \eta_{3;2'}).
  \label{eq:undici}
\end{equation}
If, on the other hand, the time-dependence of the delays is such as to
prevent the delay operators to commute, the expression of the
unequal-arm Michelson combination above no longer cancels $p_1$. In
order to derive the new expression for the unequal-arm interferometer
that accounts for ``flexing'' and simultaneously cancels the noise
from the microwave frequency signal referenced to the onboard lasers,
let us first consider the following two combinations of the one-way
measurements entering into the $X^{OFC}$ observable given in
Eq.~(\ref{eq:undici}):
\begin{eqnarray}
\left[I_1 I_{2'}(I_3 \eta_{1'} + \eta_{3;2'}) + (I_{2'} \eta_{1} + \eta_{2';3})_{;22'}\right] & = &
\left[D_{2'}D_{2}D_{3}D_{3'} - I_1I_{1'}I_{2'}I_3 \right] p_1 \ ,
\label{eq:ventinovea}
\\
\left[I_{1'}I_3 (I_{2'}\eta_{1} + \eta_{2';3}) + (I_3\eta_{1'} + \eta_{3;2'})_{;3'3}\right] & = &
\left[D_{3}D_{3'}D_{2'}D_{2} - I_1 I_{1'} I_{2'} I_3 \right] p_1 \ .
\label{eq:ventinoveb}
\end{eqnarray}

Using Equations~(\ref{eq:ventinovea}, \ref{eq:ventinoveb}) we can use
the ``delay technique'' again to finally derive the following expression
for the new unequal-arm Michelson combination $X^{OFC}_1$. This new
TDI combination accounts for the flexing effect and cancels the noise
of the microwave signal referenced to the laser frequency through the
OFC technique
\begin{eqnarray}
  X^{OFC}_1 & = & \left[D_{2'}D_{2}D_{3}D_{3'} - I_1I_{1'}I_{2'}I_3 \right]
\left[I_{1'}I_3 (I_{2'}\eta_{1} + \eta_{2';3}) + (I_3\eta_{1'} + \eta_{3;2'})_{;3'3}\right]
\nonumber \\
  & & - \left[D_{3}D_{3'}D_{2'}D_{2} - I_1 I_{1'} I_{2'} I_3 \right]
\left[I_1 I_{2'}(I_3 \eta_{1'} + \eta_{3;2'}) + (I_{2'} \eta_{1} +
  \eta_{2';3})_{;22'}\right] \ .
  \label{eq:trenta}
\end{eqnarray}
As usual, $X^{OFC}_2$ and $X^{OFC}_3$ are obtained by cyclic
permutation of the spacecraft indices. This expression is readily
shown to be laser-noise-free to first order of spacecraft separation
velocities $\dot L_i$ (it is ``flex-free'') and it simultaneously
removes the phase noise from the microwave signal (generated by the
OFC technique) used in the heterodyne measurements.

%%%%%%%%%%%%%%%%%%%%%%%%%%%%%%%%%%%%%%%%%%%%%%%%%%%%%%%%%%%%%%%%%%%%%%%%%%%%%%%%%%%
%%%%%%%%%%%%%%%%%%%%%%%%%%%%%%%%%%%%%%%%%%%%%%%%%%%%%%%%%%%%%%%%%%%%%%%%%%%%%%%%%%%

\subsection{The Sagnac combinations}
\label{sec:Sagnac}

In the case of the Sagnac variables $(\alpha, \beta, \gamma,
\zeta)$~\cite{TD2005} light originating from a spacecraft is
simultaneously sent around the array on clockwise and
counter-clockwise loops, and the two returning beams are then
recombined. If the array is rotating and flexing, the two beams
experience a different delay due to the rotation of the array (the
Sagnac effect) as well as the time-dependence of the armlengths and
the resulting Doppler effects.

Since the previously derived Sagnac TDI combinations no longer
simultaneously cancel the laser and the microwave signal noises, in
order to derive the new TDI expressions let us first write down the
six terms entering, for instance, into the $\alpha$ combination
suitable for a stationary array (see equation (60) in ~\cite{TD2005})
in the attempt to explicitly identify the expression of the new TDI
combination, $\alpha^{OFC}$. This combination must satisfy the property
of removing the same laser fluctuations affecting two beams that have
been made to propagated clockwise and counter-clockwise around the
array.
The expression for $\alpha$ (equation (60) of ~\cite{TD2005})
contains six terms, i.e. $\eta_{1'}, \D_{2'} \eta_{3'}, \D_{1'}
\D_{2'} \eta_{2'}, \eta_{1}, \D_{3} \eta_{2}, \D_{1} \D_{3} \eta_{3}$,
which now assume the following forms
\begin{eqnarray}
\eta_{1'} & = & \D_{2'} p_3 - I_{1'} p_1 \ ,
\label{alphaTerms_a}
\\
\D_{2'} \eta_{3'} & = & \D_{2'} \D_{1'} p_2 - I_{3'} \D_{2'} p_3 \ ,
\label{alphaTerms_b}
\\
\D_{1'} \D_{2'} \eta_{2'} & = & \D_{1'} \D_{2'} \D_{3'} p_1 - I_{2'} \D_{1'} \D_{2'} p_2 \ ,
\label{alphaTerms_c}
\\
\eta_{1} & = & \D_3 p_2 - I_1 p_1 \ ,
\label{alphaTerms_d}
\\
\D_{3} \eta_{2} & = & \D_{3} \D_1 p_3 - I_2 \D_3 p_2 \ ,
\label{alphaTerms_e}
\\
\D_{3} \D_{1} \eta_{3} & = & \D_{3} \D_{1} \D_2 p_1 - I_3 \D_3 \D_1 p_3 \ .
\label{alphaTerms_f}
\end{eqnarray}
By simple inspection of the above expressions it is easy to see that
the following two linear combinations of the above six measurements
only contain the laser noise $p_1$
\begin{eqnarray}
\alpha_{\uparrow} & \equiv & I_{2'}I_{3'} \eta_{1'} + I_{2'} \D_{2'} \eta_{3'} 
+ \D_{1'} \D_{2'} \eta_{2'} = [\D_{1'} \D_{2'} \D_{3'} - I_{1'}I_{2'}I_{3'}] p_1 \ ,
\label{alphaup}
\\
\alpha_{\downarrow} & \equiv & I_2 I_{3} \eta_1 + I_3 \D_{3} \eta_{2} + \D_{3}
\D_{1} \eta_{3}  = [\D_{3} \D_{1} \D_2 - I_1 I_2 I_{3}] p_1 \ .
\label{alphadown}
\end{eqnarray}
Since the delay operators do not commute, we conclude that the
straight difference $\alpha_{\uparrow} - \alpha_{\downarrow}$ does not
cancel the laser noise. However, if we now apply the ``delay technique''
~\cite{TEA04,TD2005} to the combinations given in
Eqs.~(\ref{alphaup}, \ref{alphadown}) we finally get the expression
for the Sagnac combination $\alpha^{OFC}_1$
\begin{equation}
  \alpha^{OFC}_1  \equiv [\D_{3} \D_{1} \D_2 - I_1 I_2 I_3] \alpha_{\uparrow}
  - [\D_{1'} \D_{2'} \D_{3'} - I_{1'} I_{2'} I_{3'}] \alpha_{\downarrow} \ ,
  \label{alpha1}
\end{equation}
Although the combination $\alpha^{OFC}_1$ still shows the presence of a
residual laser noise, it can be shown to be small as it involves the
difference of the clockwise and counter-clockwise rates of change of
the propagation delays on the \emph{same} circuit. For LISA, the
remaining laser phase noises in $\alpha^{OFC}_i$, $i = 1, 2, 3$, are
several orders of magnitude below the secondary noises.

In order to derive the new TDI expression for the fully symmetric
Sagnac combination, $\zeta$, we remind the reader that the rotation of
the array breaks the symmetry and therefore its uniqueness. However,
there still exist three generalized TDI laser-noise-free data
combinations that have properties very similar to $\zeta$, and which
can be used for the same scientific purposes~\cite{TAE01}. These
combinations, which we call $(\zeta^{OFC}_1, \zeta^{OFC}_2,
\zeta^{OFC}_3)$, can be derived by applying again our time-delay
operator approach.

\noindent
In order to proceed with the derivation of $\zeta^{OFC}_1$, we should
first remind ourselves that this TDI combination contains all six
$\eta_{i}$, $\eta_{i'} \ , \ i=1, 2, 3$, each of which is delayed only
once~\cite{AET99}. By simple inspection of the TDI combination
$\zeta_1$ given in Eq. (67) of reference ~\cite{TD2005}, we will start
by considering the following six terms entering in it
\begin{eqnarray}
\eta_{3,3} & = & \D_3 \D_2 p_1 - I_3 \D_3 p_3 \ ,
\nonumber
\\
\eta_{3',3} & = & \D_3 \D_{1'} p_2 - I_{3'} \D_3 p_3 \ ,
\nonumber
\\
\eta_{1,1'} & = &   \D_{1'} \D_3 p_2 - I_1 \D_{1'} p_1 \ ,
\nonumber
\\
\eta_{1',1} & = & \D_1 \D_{2'} p_3 - I_{1'} \D_1 p_1 \ ,
\nonumber
\\
\eta_{2,2'} & = & \D_{2'} \D_1 p_3 - I_2 \D_{2'} p_2 \ ,
\nonumber
\\
\eta_{2',2'} & = &  \D_{2'} \D_{3'} p_1 - I_{2'} \D_{2'} p_2 \ .
\nonumber
\\
\label{eq:ventidueb}
\end{eqnarray}
If the array is rigidly rotating (i.e. its arm lengths are
constant), it is easy to see that the following combinations contain
only the noise from laser \# 1
\begin{eqnarray}
I_3 \eta_{1,1'} - I_3 \eta_{3',3} + I_{3'} \eta_{3,3} & = & [I_{3'} \D_3
\D_2 - I_1 I_3 \D_{1'}] p_1 \ ,
\label{zetaup}
\\
I_{2'} \eta_{1',1} - I_{2'} \eta_{2,2'} + I_{2} \eta_{2',2'} & = & [I_{2} \D_{2'}
\D_{3'} - I_{1'} I_{2'} \D_{1}] p_1 \ ,
\label{zetadwn}
\end{eqnarray}
as we have used the commutativity property of the delay operators
in order to cancel the $p_2$ and $p_3$ terms. Since both sides
of the two equations above contain only the $p_1$ noise,
$\zeta^{OFC}_1$ is found by the following expression:
\begin{eqnarray}
\zeta^{OFC}_1 & = & \left[I_2 D_{2'}D_{3'} - I_{1'} I_{2'} D_{1}\right]
  \left(I_3 \eta_{1,1'} - I_3 \eta_{3',3} + I_{3'} \eta_{3,3}\right) 
\nonumber
\\
& - &  \left[I_{3'} D_{3}D_{2} - I_1 I_3 D_{1'}\right]
\left(I_{2'} \eta_{1',1} - I_{2'} \eta_{2,2'} + I_{2}
\eta_{2',2'}\right) \ ,
\label{eq:ventitre}
\end{eqnarray}
which is a generalization of Eq. (67) given in ~\cite{TD2005}. If the
delay-times also change in time, the perfect cancellation of the laser
noises is no longer achieved in the $(\zeta^{OFC}_1, \zeta^{OFC}_2,
\zeta^{OFC}_3)$ combinations derived above. However, following the
considerations made for the corresponding second-generation TDI
combinations derived in ~\cite{TEA04}, it can be shown that for a
mission like LISA the magnitude of the residual laser noises in these
combinations are significantly smaller than the secondary system
noises, making their effects entirely negligible.

\subsection{The Monitor, Beacon, and Relay Combinations}
\label{sec:others}

The expressions of the Relay ($U^{OFC}_1$), Beacon ($P^{OFC}_1$), and
Monitor ($E^{OFC}_1$) combinations that account for the rotation,
flexing of the array, and implementation of the OFC technique, can
also be derived by applying the time-delay iterative procedure
highlighted in the previous subsections. However, as also noted in
~\cite{TEA04}, in order to reduce the number of terms in the
monitor combinations we will rely on the use of the inverse delay
operator $D_i^{-1} \ , \ D_{i'}^{-1}\ , \ i = 1, 2, 3$.

\subsubsection{The relay}

In a {\it relay} combination, $U^{OFC}_1$, one spacecraft is capable
of only receiving a laser beam along one arm and transmitting along
the other. If, for instance, we take spacecraft \# 1 to be the relay
spacecraft, the resulting TDI combination may contain the following
four measurements~\cite{TEA04}: $\eta_{1'} \ , \ \eta_2 \ , \
\eta_{2'} \ , \ \eta_{3'}$. If we now consider the expressions given
in Eqs. (44-45) derived in~\cite{TEA04} for the relay combination
without OFC, it is relatively easy to obtain from them the following equivalent
expressions valid when the OFC technique is implemented
\begin{eqnarray}
I_{1'}I_{2}I_{3'} \eta_{2'} + I_{2}I_{3'} \eta_{1',3'} +
I_{2}\eta_{3';2'3'} + \eta_{2;1'2'3'} - I_{1'}I_{2'}I_{3'} \eta_{2} &
= & [\D_{3'}\D_{2'}\D_{1'} - I_{1'}I_{2'}I_{3'}]\D_{1} p_3 \ ,
\label{U11}
\\
I_{1'} \eta_{2';1'1} + I_{1'}I_{2'} \eta_{3',1} + \eta_{1';3'1'1} & =
& \D_{1}[\D_{1'}\D_{3'}\D_{2'} - I_{1'}I_{2'}I_{3'}] p_3 \ .
\label{U12}
\end{eqnarray}
By applying the operator $\D_{1}[\D_{1'}\D_{3'}\D_{2'} -
I_{1'}I_{2'}I_{3'}]$ to the left-hand-side of Eq.(\ref{U11}), the
operator $[\D_{3'}\D_{2'}\D_{1'} - I_{1'}I_{2'}I_{3'}]\D_{1}$ to the
left-hand-side of Eq. (\ref{U12}), and then take the difference of the
resulting two expressions, we get the following expression for the
second-generation relay combination when OFC is implemented
\begin{eqnarray}
U_1^{OFC} & = & \D_{1}[\D_{1'}\D_{3'}\D_{2'} - I_{1'}I_{2'}I_{3'}]
\left\{
I_{1'}I_{2}I_{3'} \eta_{2'} + I_{2}I_{3'} \eta_{1',3'} +
I_{2}\eta_{3';2'3'} + \eta_{2;1'2'3'} - I_{1'}I_{2'}I_{3'} \eta_{2} \right\}
\nonumber
\\
& - & [\D_{3'}\D_{2'}\D_{1'} - I_{1'}I_{2'}I_{3'}]\D_{1} \left\{
I_{1'} \eta_{2';1'1} + I_{1'}I_{2'} \eta_{3',1} + \eta_{1';3'1'1}
\right\}
\label{U}
\end{eqnarray}
with $U_2^{OFC} \ , \ U_3^{OFC}$ obtained by cycling the spacecraft
indeces. It can readily be verified that the laser noise remaining in
this combination vanishes to first order in the inter-spacecraft
velocities.

\subsubsection{The beacon}

In the {\it beacon} combination, say $P_1^{OFC}$, spacecraft \# 1
transmits (only) to the other two while the other two exchange laser
light as usual.  In this operating mode only the measurements
$\eta_{2} \ , \ \eta_{2'} \ , \ \eta_{3} \ , \ \eta_{3'}$ will be
available for synthesizing the combination $P_1^{OFC}$. To identify
its expression we proceed by modifying Eqs. (49-50) given in
~\cite{TEA04} in such a way to exactly cancel the $p_2$, $p_3$ laser
noises while retaining $p_1$. After some long but straightforward
algebra we get the following two expressions that contain only $p_1$
\begin{eqnarray}
I_2 I_3 \eta_{3',3'} + I_3 \eta_{2;1'3'} + \eta_{3;11'3'} - I_2I_{3'} \eta_{3,3'} 
& = & \D_{3'}[\D_{1'}\D_{1} - I_2I_{3'}]\D_2 p_1 \ ,
\label{P11}
\\
I_{2'} I_{3'} \eta_{2,2} + I_{2'} \eta_{3';12} + \eta_{2';1'12} -
I_2I_{3'} \eta_{2',2} 
& = & \D_{2}[\D_{1}\D_{1'} - I_2I_{3'}]\D_{3'} p_1 \ .
\label{P12}
\end{eqnarray}
By now applying our time-delay iterative procedure we obtain the final
expression for $P_1^{OFC}$
\begin{eqnarray}
P_1^{OFC} & = & \D_{2}[\D_{1}\D_{1'} - I_2I_{3'}]\D_{3'} [I_2 I_3 \eta_{3',3'} + I_3
\eta_{2;1'3'} + \eta_{3;11'3'} - I_2I_{3'} \eta_{3,3'}] 
\nonumber
\\
& - & \D_{3'}[\D_{1'}\D_{1} - I_2I_{3'}]\D_2 
[I_{2'} I_{3'} \eta_{2,2} + I_{2'} \eta_{3';12} + \eta_{2';1'12} -
I_2I_{3'} \eta_{2',2}] 
\label{P1}
\end{eqnarray}

\subsubsection{The monitor}

In the {\it monitor} combination, $E_1^{OFC}$ for instance, spacecraft
\# 1 receives (only) from the other two while the other two exchange
laser light as usual; $E_1^{OFC}$ therefore relies only on the
measurements $\eta_{1} \ , \ \eta_{1'} \ , \ \eta_{2} \ , \
\eta_{3'}$. If we now consider the expressions given
in Eqs. (54-57) of ~\cite{TEA04} for the monitor combination
without OFC, it is relatively easy to derive from them the following equivalent
expressions valid when the OFC technique is implemented
\begin{eqnarray}
\eta_{1;11'} - I_2 I_{3'} \eta_{1} & = & [\D_{1'}\D_{1} - I_2I_{3'}]\D_3 p_2 - I_{1} [\D_{1'}\D_{1} - I_2I_{3'}] p_1 \ ,
\label{E11}
\\
I_{3'} \eta_{2,3} + \eta_{3';13} & = & \D_3 [\D_{1}\D_{1'} -I_2I_{3'}] p_2 \ ,
\label{E12}
\\
\eta_{1';1'1} - I_2 I_{3'} \eta_{1'} & = & [\D_{1}\D_{1'} - I_2I_{3'}]\D_{2'} p_3 - I_{1'} [\D_{1}\D_{1'} - I_2I_{3'}] p_1 \ ,
\label{E13}
\\
I_{2} \eta_{3',2'} + \eta_{2;1'2'} & = & \D_{2'} [\D_{1'}\D_{1} - I_2I_{3'}] p_3 \ .
\label{E14}
\end{eqnarray}
The above expressions can be first combined in pairs to remove the
$p_2$, $p_3$ noises in the following way
\begin{eqnarray}
[\D_{1'}\D_{1} - I_2I_{3'}]\D_3 [I_{3'} \eta_{2,3} + \eta_{3';13}] 
& - & \D_3 [\D_{1}\D_{1'} -I_2I_{3'}][\eta_{1;11'} - I_2 I_{3'}
\eta_{1}] 
\nonumber
\\
& = & I_{1} \D_3 [\D_{1}\D_{1'} -I_2I_{3'}] [\D_{1'}\D_{1} - I_2I_{3'}] p_1 \ ,
\label{E15}
\end{eqnarray}
\begin{eqnarray}
[ \D_{1}\D_{1'} - I_2I_{3'}] \D_{2'}[I_{2} \eta_{3',2'} + \eta_{2;1'2'}]
& - & \D_{2'} [ \D_{1'}\D_{1} - I_2I_{3'}][\eta_{1';1'1} - I_2 I_{3'}
\eta_{1'}] 
\nonumber
\\
& = &
I_{1'} \D_{2'} [ \D_{1'}\D_{1} - I_2I_{3'}][ \D_{1}\D_{1'} - I_2I_{3'}]
p_1 \ .
\label{E16}
\end{eqnarray}
To find the final expression for $E_1^{OFC}$ we could apply again the
iterative procedure. However, to get an expression that has a smaller
number of terms ~\cite{TEA04}, we will use the inverse delay
operators, $\D_{2'}^{-1}$, $\D_{3}^{-1}$. By simply inspecting the
right-hand-sides of Eqs. (\ref{E15}, \ref{E16}), it is easy to derive
the following expression for $E_1^{OFC}$
\begin{eqnarray}
E_1^{OFC} & & = 
I_{1'} \D_{3}^{-1}[\D_{1'}\D_{1} - I_2I_{3'}]\D_3 [I_{3'} \eta_{2,3} +
\eta_{3';13}] 
- I_{1'}[\D_{1}\D_{1'} -I_2I_{3'}][\eta_{1;11'} - I_2 I_{3'}
\eta_{1}] 
\nonumber
\\
& - & I_{1} \D_{2'}^{-1} [ \D_{1}\D_{1'} - I_2I_{3'}] \D_{2'}[I_{2}
\eta_{3',2'} + \eta_{2;1'2'}]
+ I_{1} [ \D_{1'}\D_{1} - I_2I_{3'}][\eta_{1';1'1} - I_2 I_{3'}
\eta_{1'}] \ ,
\label{E1}
\end{eqnarray}
which can easily be shown to be laser noise-free to first order in the
systematic relative velocities of the spacecraft.

\section{Summary and Conclusions}
\label{SECIV}

We have derived second-generation time-delay interferometric
combinations valid when the microwave signal used for heterodyning the
phase measurements is generated by an onboard optical-frequency comb
subsystem rather than a USO. This provides a microwave signal that is
coherent to the frequency of the onboard stabilized laser, and results
in a significant simplification of the onboard interferometry
system. This is because (i) generation of modulated beams and
additional heterodyne measurements involving sidebands are no longer
needed, and (ii) the entire onboard USO subsystem can be replaced with
the microwave signal referenced to the onboard laser.  The
corresponding hardware simplification results in a considerably
reduced system complexity and probability of subsystem failure. Recent
progress in the realization of a space-qualified OFC indicates that
such a capability will be available before the planned flight of the
ESA gravitational wave mission eLISA ~\cite{eLISA}.

\section*{Acknowledgments}

We acknowledge financial support provided by the Jet Propulsion
Laboratory Research \& Technology Development program. This research
was performed at the Jet Propulsion Laboratory, California Institute
of Technology, under contract with the National Aeronautics and Space
Administration.

\end{document}